# Effect of cobalt substitution on structural, impedance, ferroelectric and magnetic properties of multiferroic $Bi_2Fe_4O_9$ ceramics

S. R. Mohapatra[1], B. Sahu[1], M. Chandrasekhar[1], P. Kumar[1], S. D. Kaushik[2], S. Rath[3], A. K. Singh[1*]

[1] Department of Physics and Astronomy, National Institute of Technology Rourkela-769008, Odisha, India

[2]UGC-DAE Consortium for Scientific Research Mumbai Centre, R-5 Shed, BARC, Mumbai-400085, India

[3]School of Basic Sciences, Indian Institute of Technology Bhubaneswar-751007, Odisha, India

**Abstract:** Structural, impedance, ferroelectric and magnetic properties were examined in multiferroic $Bi_2Fe_{4(1-x)}Co_{4x}O_9$ ($0 \leq x \leq 0.02$) ceramics synthesized via solid-state reaction method. X-ray diffraction analysis and Rietveld refinement showed secondary phase formation (for $x \geq 0.01$) which was subsequently confirmed from room temperature Raman spectroscopy study. The frequency dependence of impedance and electric modulus of the material showed the presence of non-Debye type relaxation in all the samples. The values of the activation energies calculated from imaginary impedance and modulus lie in the range of 0.92-0.99 eV which confirmed that the oxygen vacancies play an important role in the conduction mechanism. Moreover, suitable amount of Co substitution significantly enhanced the remnant polarisation ($2P_r$) from 0.1193 μC/cm² (x = 0) to 0.2776 μC/cm² (x = 0.02). Besides, room temperature M-H measurement showed improved ferromagnetic hysteresis loop for all the modified samples. The remnant magnetization ($M_r$) and coercive field ($H_c$) increased from 0.0007 emu/gm and 42 Oe for x = 0 to 0.1401 emu/gm and 296 Oe for x = 0.02. The improved ferroelectricity was due to Co $3d$-O $2p$ hybridization and enhanced magnetization originated from the partial substitution of $Co^{3+}$ ions leading to breakdown of balance between the anti-parallel sub lattice magnetization of $Fe^{3+}$ ions.

*Email: singhanil@nitrkl.ac.in, Phone: +91-661-2462731; Fax: +91-661-2462739

## 1. Introduction

Multiferroics unveils the simultaneous existence of two or more ferroic orders such as ferroelectricity (FE) and ferromagnetism (FM) with rich physics involved in it [1,2]. The coexistence of several order parameters brings out novel physical phenomena and offers possibilities for new device functions. Multiferroic materials are enriched with lots of potential applications such as sensors, filters, data storage, multiple-state memories and devices for spintronics [3-6]. The coupling interaction between various ferroic orders result in novel physical phenomenon such as magnetoelectric (ME) effect, which is of great interest due to its importance in basic research and applications [7,8]. Among these multiferroics, the $Bi_2M_4O_9$ (M = $Al^{3+}$, $Ga^{3+}$, $In^{3+}$ and $Fe^{3+}$) compounds with a mullite-type structure have attracted significant attention due to its wide scale of applications in solid state oxide fuel cells, electro-chemical reactors, sensors, gas separation membranes, scintillation materials, catalysts etc [9-11]. *K.L.Da Silva* et al. [12] reported that these bismuth-bearing oxides are orthorhombic and crystalizes in space group *'Pbam'*. $Bi_2M_4O_9$ structures are characterized by chains of edge sharing $MO_6$ octahedra which are interconnected by $M_2O_7$ double tetrahedral units and $BiO_6$ 'E' groups (where 'E' refers to a $6s^2$ lone electron pair) in an alternating ordered manner along the c-axis. Recently, *Singh* et al. [13] has reported that $Bi_2Fe_4O_9$ (BFO) is usually formed as the solid state bi-product of $BiFeO_3$ displaying ferroelectric hysteresis loops at T = 250 K and anti-ferromagnetic ordering at $T_N$ = 260 K which are resulted due to geometrically spin frustrated anti-ferromagnetic order in the compound. BFO displays various interesting functional applications such as photo-catalyst, semi-conductor gas sensors, high performance catalysts, dielectric, magnetic and multiferroic properties [14-17]. *Ressouche* et al. has reported that a unit cell of $Bi_2Fe_4O_9$ consists of two kinds of Fe atoms (Fe1, Fe2); the first kind of Fe atom is (Fe1) enclosed by four $O^{2-}$ ions in a tetrahedral manner having an antiferromagnetic interaction among themselves whereas the second kind of Fe atom (Fe2) is octahedrally coordinated by six $O^{2-}$ ions interacting in a ferromagnetic manner. Also, $Bi^{3+}$ ions are surrounded by eight $O^{2-}$ ions held mutually orthogonal to each other with shorter as well as longer $BiO_3$ units [18, 19].

Despite being a well-known multiferroic material, $Bi_2Fe_4O_9$ possesses weak macroscopic magnetism and poor FE property at room temperature which has been a major drawback as compared to $BiFeO_3$. Hence, to achieve room temperature FM as well as improved FE property, chemical substitution in $Bi_2Fe_4O_9$ is desirable. Till now various substitution has been done at the A-site and B-site of $Bi_2Fe_4O_9$ to improve its multiferroic properties [20-22]. To the best of our knowledge, there has been no report regarding cobalt substitution in order to address the above mentioned problems. Also, substitutional effect of cobalt on the dielectric and impedance properties of $Bi_2Fe_4O_9$ is still unknown. Thus, we have selected cobalt ions as dopants because $Co^{3+}$ ions have similar ionic radius as that of $Fe^{3+}$ (0.61 Å for $Co^{3+}$ and 0.645 Å for $Fe^{3+}$ both with octahedral coordination), $Co^{3+}$ is a transition metal ions and it improves the magnetic properties as reported in literatures [23, 24]. Hence, we report the preparation of $Co^{3+}$ substituted $Bi_2Fe_4O_9$ via solid state route and explored their structural properties by combined X-ray diffraction, Rietveld refinement, field emission scanning electron microscopy and RAMAN spectroscopy. In parallel, dielectric study followed by complex impedance study were carried out to investigate and analyse the electrical processes occurring in the material. Lastly, it is interesting to note that slight substitution of Co plays an important role in enhancing the FE as well as FM properties in the $Bi_2Fe_4O_9$ modified systems.

## 2. Experimental details

The polycrystalline samples of $Bi_2Fe_{4(1-x)}Co_{4x}O_9$ (x = 0, 0.005, 0.01, 0.015 and 0.02) were prepared using high purity (>99.9%, Sigma Aldrich) oxides namely $Bi_2O_3$, $Fe_2O_3$ and $Co_3O_4$ by solid state reaction route. The above chemicals were thoroughly mixed in stoichiometric proportions, grounded for 2 hours and calcined at 800 °C for 12 hours followed by a sintering at 850 °C for 10 hours. The phase formation of synthesized compounds was verified by room temperature X-ray diffraction (XRD) carried out using multipurpose X-ray diffraction system (RIGAKU, JAPAN). XRD was carried out with Cu-Kα radiation (λ = 1.54 Å) in a wide range of Bragg angles ($10° \leq 2\theta \leq 70°$) with a step size of 0.002° at a slow scan rate of 3°/min. X-ray photoemission spectroscopy (XPS) is performed using Al-Kα (hʋ =1486.6 eV) lab-source. The surface morphology was investigated using Field Emission Scanning Electron Microscope (FESEM). Raman spectroscopy was measured at room temperature using a laser Raman spectrometer (Horribba Scientific Instruments T6400) with monochromatic wavelength 514 nm. The power of the incident laser beam was 20 mW. Dielectric and impedance measurements were performed using High precession impedance analyzer (Wayne Kerr 6500B) for a temperature range of 50 - 360 °C over a wide frequency range (0.1-1000 kHz). Electrodes were made by applying silver paste on both sides of the samples and then fired at 250 °C for 1 hour in order to use it as a parallel plate capacitor. The ferroelectric properties of the ceramics were measured on a standard ferroelectric tester using a Radiant precision premier II. Finally, magnetization study was carried out using commercial 9T re-liquefier based Physical Property Measurement System (PPMS)-Vibrating Sample Magnetometer (Quantum Design, USA).

## 3. Results and Discussion

### 3.1. X-ray diffraction and microstructural analysis

Fig. 1 (a) shows the room temperature powder XRD patterns of $Bi_2Fe_{4(1-x)}Co_{4x}O_9$ (x = 0, 0.005, 0.01, 0.015 and 0.02) abbreviated as BFO, BFCO0.5, BFCO1, BFCO1.5 and BFCO2 respectively. All the major peaks are identified and indexed which are consistent with the standard data (JCPDS no. 25-0090). XRD pattern of BFO, BFCO0.5 indicates single phase formation (orthorhombic phase) with space group *'Pbam'* whereas cobalt doped BFO (x ≥ 0.01) indicates the presence of a secondary phase i.e., $BiFeO_3$ at $2\theta \sim 32^0$ (rhombohedral phase with *'R3c'* space group). The Rietveld refinement was obtained using *FULLPROF* software [25] and the refinement of BFO, BFCO1 and BFCO2 is shown in fig 1 (b). The amount of secondary phase is well within the 5% range for BFCO1 to BFCO2. The obtained fitting parameters from structural refinement are presented in table 1 and indicate a good agreement between the observed and refined patterns. However, we observe decrease in lattice parameters with increasing x concentration along with the shifting of main peaks i.e., (002) and (220) to higher angles, as can be seen from the inset of fig. 1(a). The above feature can be explained because of the lower ionic radius of $Co^{3+}$ ion substituting the higher ionic radii of $Fe^{3+}$ ion. The shifting of the (002) and (220) peak towards higher angle ($2\theta$) nullifies the possibility of $Co^{2+}$ ion substitution at $Fe^{3+}$ site which is verified with XPS study. The XPS analysis helps in understanding the oxidation states of the ions present in the samples. Fig 1 (c) illustrates the deconvolution of Co-2p spectrum of BFCO2 sample indicating the existence of two main peaks: (i) $Co-2p_{3/2}$ at 776.6 eV of $Co^{2+}$ and 780.7 eV of $Co^{3+}$ and (ii) $Co-2p_{1/2}$ at 796.1 eV of $Co^{2+}$ and 801.5 eV of $Co^{3+}$ respectively; along with two shake-up satellite peaks at 789.1 eV and 792.8 eV [26]. As seen from fig 1 (c), the presence of $Co^{3+}$ ion is ~85% whereas $Co^{2+}$ ion is ~15% indicating the existence of $Co^{3+}$ ion as the major contribution in our Co substituted BFO ceramics. Thus, XPS analysis is in good agreement with our XRD data.

The FESEM images of sintered ceramics (BFO and BFCO2) are shown in fig. 2. The microstructures consist of well developed, non-uniform large grains, in-homogenously distributed on the surface of the samples. Due to Co substitution, average grain size is found to increase from ~2μm for BFO to ~4μm for BFCO2. This increase in grain size may be attributed to the oxygen vacancies, since motion of oxygen vacancies during sintering promotes grain growth [19]. It can also be seen that the sample exhibits relatively dense grain with less degree of porosity with increase in x content.

### 3.2. Raman study

Raman spectroscopy is a powerful tool to probe structural as well as vibrational properties of a material. To have better understanding regarding the structural changes observed in XRD, room temperature Raman spectra of BFO and Co modified BFO samples were performed and shown in fig. 3. *Iliev* et al. have estimated 42 Raman modes ($12A_g + 12B_{1g} + 9B_{2g} + 9B_{3g}$) for orthorhombic $Bi_2Fe_4O_9$ using group theoretical irreducible representations.[27] However, we observe less no of modes as compared to group theoretical calculations due to the insufficient intensity arising from small polarizability of several modes. The assigned Raman modes are in good agreement with previous reported data [24,27]. *Hermet* et al. suggested that Bi atoms are assigned to low frequency modes (up to 167 $cm^{-1}$) and Fe atoms actively participate in the modes between 152 and 262 $cm^{-1}$, while for higher frequency modes (above 262 $cm^{-1}$), oxygen motion strongly dominates [28]. The Raman modes observed for Co doped $Bi_2Fe_4O_9$ are listed in table 2. From fig. 3, we found that there is no significant change in the Raman modes between BFO and Co modified BFO, yet small variation in the spectra is observed. Due to Co substitution at Fe sites, we noticed the shifting of Raman modes towards lower wavenumber. This could be attributed to (i) lower atomic mass of Fe (55.845 amu) as compared to Co (58.933 amu) and (ii) lattice disorder. This relative shift of various Raman modes with increasing x content specifies that properties of BFO are affected due to substitutional action of the dopant which we shall discuss further in other characterizations. Intense Raman mode formed at 220 $cm^{-1}$ (BFO), 219 $cm^{-1}$ (BFCO0.5), 209 $cm^{-1}$ (BFCO1) and 196 $cm^{-1}$ (BFCO2) is related to Fe-O vibration in $FeO_4$ tetrahedron. Similarly, other phonon modes observed for BFO (285 $cm^{-1}$, 317 $cm^{-1}$ and 424 $cm^{-1}$), BFCO0.5 (284 $cm^{-1}$, 318 $cm^{-1}$ and 426 $cm^{-1}$), BFCO1 (279 $cm^{-1}$, 320 $cm^{-1}$ and 427 $cm^{-1}$), BFCO1.5 (275 $cm^{-1}$, 321 $cm^{-1}$ and 427 $cm^{-1}$) and BFCO2 (274 $cm^{-1}$, 324 $cm^{-1}$ and 428 $cm^{-1}$) owes to vibration of Fe-O in the $FeO_6$ octahedron [29]. It is also important to notice an emergence of additional peak for compositions $x \geq 0.01$ at 558 $cm^{-1}$ (BFCO1), 561 $cm^{-1}$ (BFCO1.5) and 564 $cm^{-1}$ (BFCO2) as shown in fig. 3 (marked in dotted circle) which indicates the overlapping of the torsional mode related to the $FeO_6$ and $CoO_6$ octahedra. Further, the above band is observed to be blue shifted with increase in Co dopant concentration which may be due to the stabilization of $FeO_6$ octahedron [30]. This peak can be prominently seen for BFCO2 (564 $cm^{-1}$) as depicted in the inset of fig. 3. Moreover, inset of fig 3 shows the room temperature deconvolated Raman spectra of BFCO2 in order to locate various Raman modes which is fitted to the sum of seven Lorenzian peaks. The weak mode around at 700 $cm^{-1}$ refers to torsional bending of $FeO_6$ and $CoO_6$ octahedron [30]. These modes represent active modes of BiFeO₃ which further confirms the onset of secondary phase thus supporting our XRD data as well as previously reported study [24].

### 3.3. Frequency dependant dielectric study

Fig. 4 (a-c) shows the variation of frequency dependant dielectric permittivity (ε') and inset of fig. 4(a-c) shows tan loss for the BFO, BFCO1 and BFCO2 ceramic at different temperatures ranging from 50 °C - 360 °C. It is clear from fig 4 (a-c) that $\varepsilon'$ decreases smoothly in both parent and substituted samples. This indicates dielectric relaxation phenomenon to be associated with a frequency dependent orientational polarization [31]. Generally, frequency dependent complex dielectric permittivity ($\varepsilon^*$) can be well defined by the modified Cole-Cole relaxation equation [32] described below:

$$\varepsilon^*(\omega) = \varepsilon_\infty + \frac{(\varepsilon_s - \varepsilon_\infty)}{[1 + (i\omega\tau)^{1-\alpha}]} \tag{1}$$

In the low frequency range (<10 kHz), ε' increases with increasing temperature and decreases with increase in frequency for all the samples. On the other hand, at high frequency ($\omega >> 1/\tau$), ε' becomes negligible as dipoles can no longer follow the field. Thus, magnitude of polarisation drops as frequency is increased [33]. A decrease in the values of real part of dielectric constant is observed with Co concentration which indicates substitution has a considerable effect on the dielectric response in BFO modified ceramics. It is clear from fig. 4 (a-c) that even a small quantity of x concentration can significantly change the dielectric constant of $Bi_2Fe_4O_9$. For example, the dielectric constant of BFCO2 ($\varepsilon' \sim 260$) measured at 1 kHz was found to be nearly ten times smaller than that of BFO ($\varepsilon' \sim 2690$). Due to lower ionic radius of Co ions, the substitution of Fe with Co at B-site reduces vibration space in oxygen octahedron and this would lead to decrease in dielectric polarization. For all the samples, we observe high dielectric loss (inset of fig. 4(a-c)) at lower frequency (<10 kHz) which is due to the accumulation of free charges at the interface (space-charge polarisation). Conversely, at higher frequency charge accumulation decreases, as a result, value of tan loss also decreases. The dielectric loss in doped samples increases which can be understood in terms of the space-charge-limited conduction associated with crystal defects. In tan loss we could see relaxation in BFO arising above 240 ⁰C but no clear relaxation peaks are observed in BFCO1 and BFCO2 samples and the value is found to rise at low frequencies. Similar kind of result was also studied by *P. Pandit* et al. in $Bi_{1-x}La_xFeO_3$ ceramics [34]. The relaxation peaks in loss tangent curves of pure BFO ceramics at low frequencies arise due to the hopping motion of the double ionized oxygen ion vacancies among the potential barriers. The role of the dielectric constant from the oxygen vacancy induced dielectric relaxation is considerably larger at high temperatures since, oxygen vacancy induced dielectric relaxation is basically a thermally activated process depending exponentially on temperature [34,35].

## 3.4. Impedance and Modulus spectroscopy study

Complex impedance spectroscopy (CIS) is a non-destructive tool which reveals various transport mechanisms taking place within the microstructure of the material. In general, the complex impedance ($Z^*$) is expressed as [36]:

$$Z^* = Z' - jZ'' = \frac{R_g}{(1 + j\omega R_g C_g)} + \frac{R_{gb}}{(1 + j\omega R_{gb} C_{gb})} \tag{2}$$

The variation of imaginary part of impedance (Z'') with frequency at different selected temperatures for all the samples is shown in fig. 5 (a-c). The solid lines are fitting to the obtained experimental data. Fitting is done using the equivalent circuit modelled by the Cole-Cole function [32,37] expressed as:

$$Z^* = \frac{R}{[1 + (j\omega\tau)^\alpha]} \tag{3}$$

where, $\tau = RC$ and $0 \leq \alpha < 1$ implies that the process is governed by a distribution of relaxation times. For an ideal Debye relaxation $\alpha = 1$. It is evident from fig. 5 that the experimental curves are remarkably well fitted using eqn. 3 in the entire range of frequencies and temperatures under study. The fitted value of $\alpha$ lies in the range of 0.7-0.8 for all the studied samples. The Z''-frequency patterns for all the samples displayed some characteristic features such as (i) appearance of a peak at a particular frequency (known as relaxation frequency), (ii) decrease in the magnitude of Z'' with a shift in the peak frequency toward higher frequency and (iii) peak broadening with a rise in temperature. With the increase in frequency and temperature the peak heights are found to decrease gradually and finally they merge towards the high frequency domain. It indicates the presence of space charge polarization at lower frequencies which disappears at higher frequencies [38]. The shift in peak frequency is due to the presence of electrical relaxation in the materials which relates to temperature dependent relaxation. In other words, asymmetric broadening of peaks indicates a distribution of relaxation time (τ). This feature is also reflected from the non-Debye type relaxation in the system as observed in all the cases as the centre of peak at all temperature does not lie at the same frequency. As the concentration of Co increases, the increase in the broadening and asymmetry of the peaks suggests that there is an increase in the distribution of relaxation times and departure from ideal Debye-like behaviour. This result also displays the significant effect of Co substitution on the electrical behaviour of BFO. Fig. 6 shows the complex impedance spectrum (Nyquist plot) for BFO and BFCO2 (inset) ceramic measured at temperature range 200°C – 300 °C. The radius of the Cole-Cole semicircle which measures the resistance of the material is found to decreases with increasing temperature for both BFO and BFCO2. All the semicircles exhibit some degree of depression instead of a semicircle centred at the real axis of $Z'$ due to a distribution of relaxation time. This also implies a non-Debye type relaxation mechanism occurring within the materials.

The complex electric modulus ($M^*$) study is a vital tool in determining electrical phenomenon in a dielectric system with smallest capacitance. The complex electric modulus is expressed as the reciprocal of complex dielectric permittivity ($\varepsilon^*$), given as [39]:

$$M^*(\omega) = (\varepsilon^*)^{-1} = M'(\omega) + jM''(\omega) \qquad (4)$$

where, $M'(\omega)$ and $M''(\omega)$ are real and imaginary component of complex modulus. Frequency and temperature dependence of imaginary part of complex modulus ($M''$) for BFO, BFCO1 and BFCO2 is shown in fig. 7 (a-c) respectively. We observe a single peak at each temperature in the given frequency window for all the compositions which matches very well with features discussed for $Z''$ versus f (fig. 5). The appearance of shift in $M''_{max}$ towards higher relaxation frequency with increase in temperature confirms relaxation process to be a thermally activated process where hopping mechanism of charge carriers dominates mostly at higher temperature. Also, charge carriers till the $M''_{max}$ are mobile over long distance but above peak maximum the charge carriers are mobile over short distances as they get restricted to the potential barriers [19]. The peak is found to be positioned at around the centre of the dispersion region of $M'(\omega)$ (not shown here). The frequency range corresponding to the occurrence of peak in $M''(\omega)$ indicates the transition from long range to short range mobility of charge carriers. Moreover, all the ceramic compounds under study have been detected with decrease in height of the modulus peak with an increase in Co concentration. The dielectric relaxation process in general can be represented using Laplace Transform of Kohlrausch-Williams-Watts (KWW) decay function $\phi = \phi_0 \, exp[(-t/\tau)^\beta]$; where $\tau$ is the characteristic relaxation time and $\beta$ $(0 < \beta < 1)$ is the Kohlrausch stretched exponent, which decides whether the relaxation is Debye or non-Debye in nature [40]. In order to have direct analysis of the result, Bergman modified KWW function [41,42] which is given as:

$$M''(\omega) = \frac{M''_{max}}{\left[(1-\beta) + \left(\dfrac{\beta}{1+\beta}\right)\left\{\beta\left(\dfrac{\omega_{max}}{\omega}\right) + \left(\dfrac{\omega}{\omega_{max}}\right)^\beta\right\}\right]} \qquad (5)$$

where, $M''_{max}$ is peak maximum of $M''(\omega)$, exponent $\beta$ symbolises the degree of non-Debye behaviour and is related to full width half maxima of $M''(\omega)$ vs. frequency curve. Inset of fig. 7 (a-c) shows the variation of $\beta$ with temperature for all the samples. $\beta = 1$ relates to ideal Debye nature with unique relaxation time whereas $\beta = 0$ denotes maximum interaction of dipoles with other dipoles. Also, the relaxation nature of modulus spectroscopic curve for all the samples suggest the relaxation process to be thermally activated process. However, due to high temperature synthesis of samples at 850 °C, there might be the presence of some amount of oxygen vacancy. If vacancies are created this successively helps in short range hopping of ions thus giving rise to strong relaxation [43].

The relaxation process discussed above in case of imaginary impedance and modulus suggests the relaxation process is characterized by Arrhenius type behaviour expressed as:

$$\tau = \tau_0 \exp(-E_a / k_B T) \tag{6}$$

where, $\tau_0$ is the pre-factor and $E_a$ is the activation energy for the relaxation. Fig. 8 shows the dependence of the relaxation time ($\tau$) on the reciprocal temperatures. Activation energy calculated is shown in table 3. It is known that such high values of activation energy is responsible for the migration of oxygen vacancies and suggests a possibility that the conduction in the higher temperature range for ionic charge carriers may be due to the oxygen vacancies. In our present study, we observe an enhancement in $E_a$ due to increase in x concentration. *Tamilselvan* et al. have reported that the increase in $E_a$ is related to the suppression of oxygen vacancies and vice-versa [44]. So the enhancement in $E_a$ can be explained due to the increase in resistance of our BFO modified ceramics (as seen from Nyquist plot). Additionally, the higher values of $E_a$ can be assigned with the intrinsic grain conduction which has the dominant role within our selected frequency and temperature range. The grain boundary or electrode-ceramic-interface contributions are found to be absent or may have negligible effect on the conduction process.

### 3.5. Ferroelectric properties

Fig. 9 illustrates the room temperature P-E hysteresis loops of BFO, BFCO0.5, BFCO1, BFCO1.5 and BFCO2 ceramics. All the samples exhibits clear hysteric behaviour signifying ferroelectric nature however; the hysteresis loops are still unsaturated, which implies a partial reversal of the polarization [21]. Under maximum applied electric field of 30kV/cm, the saturated polarization ($P_s$), remnant polarization ($2P_r$) and coercive electric field ($2E_c$) are found to be 0.2486 $\mu C/cm^2$, 0.1193 $\mu C/cm^2$ and 11.451 kV/cm for BFO which later increases to 0.3386 $\mu C/cm^2$, 0.2776 $\mu C/cm^2$ and 20.697 kV/cm for BFCO2. The values of $P_s$, $2P_r$ and $2E_c$ obtained for all the samples are listed in table 3. It is well understood from the figure that Co substitution has a significant effect on enhancing the ferroelectric polarisation. In addition, this enhanced ferroelectricity could also be attributed to the structural distortion induced by substitution. Similar behaviour was also observed by *Tian* et al. where it is reported that in Bi based multiferroics, a lone $s2$ pair of electrons of $Bi^{3+}$ ions hybridizes with $2p$-orbital of $O^{2-}$ ions forming a localised lobe thus leading to the non-centrosymmetric distortions and hence causes ferroelectricity [21]. In our case, in addition to $Bi^{3+}$ - $O^{2-}$ $2p$ hybridization, the presence of Co $3d$-O $2p$ hybridization stabilizes the ferroelectric distortion, which may be responsible for the enhanced ferroelectric property.

### 3.6. Magnetic properties

The magnetization hysteresis (M-H) loops of BFO modified ceramics recorded at room temperature is shown in fig. 10. Inset (a) shows the enlarged view of M-H plot in the range of -0.8–0.8 kOe for better understanding. Our BFO sample exhibits a linear magnetic field dependence of the magnetization which indicates paramagnetic nature of the material. However, deviation from the linearity to a clear hysteresis loop causing enhanced ferromagnetism is observed with increasing Co concentration. This can be due to the crossover of paramagnetic to a weak ferromagnetic state. It can be clearly seen from the figure that all the samples are not saturated even for a maximum magnetic field of 90 kOe. Table 3 shows the values of remnant magnetization ($M_r$) and coercivity ($H_c$) of the BFO and doped samples. With increase in Co substitution, $M_r$ gets monotonously enhanced from 0.00078 emu/gm for BFO to 0.1401 emu/gm for BFCO2 (~180 times) which is much higher than any other substituted ions as reported in literatures [21,22,45]. This enhanced ferromagnetic nature of the Co-doped samples may be attributed to the following possibilities: Firstly, Co substitution induces structure distortion in BFO (as seen from XRD study) thus, initiating a distorted crystal field on iron ions generating spin-orbit coupling via Dzyaloshinskii-Moriya (D-M) interactions [46]. Secondly, different magnetic moment of Co and Fe ions suppresses canted spin structure resulting in increase of magnetization. Thirdly, metal ion substitution with different valence and radius of ions may lead to breakdown of balance between the anti-parallel sub-lattice magnetization of $Fe^{3+}$ ions. Further to analyse the presence of ferromagnetic ordering in doped samples, Arrott plot ($M^2$ versus H/M) is plotted for all the compositions as shown in inset (b) of fig. 10. For BFO a steady non-linear increase in the curve is noticed whereas the S-shaped $M^2$ vs H/M are observed for x ≥ 0.005 samples. This plot shows the onset of weak ferromagnetic behaviour due to Co substitution. The linear extrapolation of the high field region of $M^2$ vs H/M plot of substituted samples yields a negative intercept on the $M^2$ axis.

This specifies the crossover from paramagnetic to weak ferromagnetic ordering [47]. As a consequence; there inevitably exists a thermal hysteresis for all the BFO modified systems as discussed above in fig. 10.

## 4. Conclusions

In conclusion, we successfully synthesized $Bi_2Fe_{4(1-x)}Co_{4x}O_9$ (x = 0, 0.005, 0.01, 0.015 and 0.02) ceramic by solid state reaction technique. Room temperature XRD and Rietveld analysis revealed the signature of secondary phase in the Co substituted BFO ceramics (for x $\geq$ 0.01) which was later confirmed by Raman spectroscopy. Impedance study reflected non-Debye type relaxation confirming significant effect of Co substitution in the BFO modified ceramics. Activation energy calculated from Z" and M" suggested conduction mechanism in the samples are governed by the thermal motion of oxygen vacancies. The improved ferroelectric and magnetic properties in the modified ceramics was due to Co *3d*-O *2p* hybridization and partial substitution of $Co^{3+}$ ions leading to breakdown of balance between the anti-parallel sub lattice magnetization of $Fe^{3+}$ ions respectively. Hence, the simultaneous presence of ferroelectric as well as magnetic hysteresis loops at room temperature makes the above studied ceramics a potential room-temperature multiferroic material.

## Acknowledgement

AKS acknowledges Board of Research in Nuclear Science (BRNS), Mumbai (Sanction No: 2012/37P/40/BRNS/2145), UGC-DAE-CSR Mumbai (Sanction No: CRS-M-187,225) and Department of Science and Technology (DST), New Delhi (Sanction No: SR/FTP/PS-187/2011) for funding. Lastly, SRM is thankful Mr. Tapabrata Dam and Mr. Rakesh Muduli for their useful suggestions.

**Figure Caption**

Fig.1 (a) Room temperature XRD pattern of polycrystalline $Bi_2Fe_{4(1-x)}Co_{4x}O_9$ ($0 \leq x \leq 0.02$) samples. Inset shows the enlarged view of main peaks [(002) and (220)]. (b) Rietveld refinement of BFO, BFCO1 and BFCO2 samples. (c) XPS spectrum of the Co-2p spectra of BFCO2 sample.

Fig.2 FESEM images of BFO and BFCO2 samples.

Fig.3 Raman spectra of $Bi_2Fe_{4(1-x)}Co_{4x}O_9$ ($0 \leq x \leq 0.02$) recorded at room temperature. Inset depicts deconvoluted Raman spectrum of BFCO2 sample fitted to the sum of seven Lorenzian peaks.

Fig.4 Frequency dependant dielectric permittivity ($\varepsilon'$) and tan loss (insets) for (a) BFO, (b) BFCO1 and (c) BFCO2 samples at various temperatures ranging from 50 $^\circ$C to 360 $^\circ$C.

Fig.5 Imaginary impedance spectrum (Z") as a function of frequency at selected temperatures for (a) BFO, (b) BFCO1 and (c) BFCO2 samples. Solid lines shows fitting using equation 3. Inset (a-c) shows enlarged view of the dotted rectangular portions.

Fig.6 Nyquist plot for (a) BFO and (b) BFCO2 samples at selected temperatures (200 $^\circ$C – 300 $^\circ$C).

Fig.7 Imaginary modulus (M") as a function of frequency at selected temperatures for (a) BFO, (b) BFCO1 and (c) BFCO2 samples. Solid lines shows the fitting using equation 5. Inset (a-c) depicts the variation of β with temperature.

Fig.8 The variation of lnτ vs 1000/T (Arrhenius plot) from imaginary impedance (Z") and modulus (M") spectrum respectively for BFO and Co modified BFO samples.

Fig.9 Room temperature polarisation hysteresis loops of $Bi_2Fe_{4(1-x)}Co_{4x}O_9$ ($0 \leq x \leq 0.02$) samples.

Fig.10 Room temperature magnetization hysteresis loops (M-H) of $Bi_2Fe_{4(1-x)}Co_{4x}O_9$ ($0 \leq x \leq 0.02$) samples. Inset (a) shows the enlarged view of M-H plot in the range of -0.8 - 0.8 kOe, (b) shows the arrot plot of BFO and Co modified BFO samples.

**Tables**

Table 1: Variation of crystal structure parameters of $Bi_2Fe_{4(1-x)}Co_{4x}O_9$ (with x = 0, 0.005, 0.01, 0.015 and 0.02)

| Samples | BFO | BFCO0.5 | BFCO1 | BFCO1.5 | BFCO2 |
|---|---|---|---|---|---|
| Space group | *Pbam* | *Pbam* | *Pbam + R3c* | *Pbam + R3c* | *Pbam + R3c* |
| Lattice parameters | | | | | |
| a (Å) | 7.97350(2) | 7.97318(5) | 7.96806(4) | 7.95993(2) | 7.95262(3) |
| b (Å) | 8.44142(4) | 8.44113(2) | 8.43627(6) | 8.43008(6) | 8.42397(6) |
| c (Å) | 6.00277(5) | 6.00271(2) | 5.99891(2) | 5.99028(5) | 5.99002(2) |
| Cell volume (Å³) | 404.032(4) | 403.998(7) | 403.250(1) | 401.964(5) | 401.287(3) |
| R-factors (%) | | | | | |
| Rp | 13.2 | 12.3 | 11.5 | 13.1 | 12.7 |
| Rwp | 15.1 | 14.4 | 14.2 | 14.9 | 15.3 |
| $\chi^2$ | 3.66 | 2.94 | 2.78 | 3.02 | 2.96 |

Table 2: Various active Raman modes observed for BFO modified samples recorded at room temperature.

| Active Raman modes [27] → Samples ↓ | $B_{2g}(2)$ | $A_g(3),A_g(4),$ $B_{1g}(3),B_{2g}(3),$ $B_{3g}(3)$ | $B_{1g}(4)$ | $A_g(5)$ | $A_g(8),B_{1g}(7),$ $B_{2g}(7),B_{3g}(7)$ | $A_g(9),B_{1g}(8),$ $B_{3g}(8)$ |
|---|---|---|---|---|---|---|
| BFO | 162 cm⁻¹ | 220 cm⁻¹ | 285 cm⁻¹ | 317 cm⁻¹ | 424 cm⁻¹ | - |
| BFCO0.5 | 161 cm⁻¹ | 219 cm⁻¹ | 284 cm⁻¹ | 318 cm⁻¹ | 426 cm⁻¹ | - |
| BFCO1 | 160 cm⁻¹ | 209 cm⁻¹ | 279 cm⁻¹ | 320 cm⁻¹ | 427 cm⁻¹ | 558 cm⁻¹ |
| BFCO1.5 | 158 cm⁻¹ | 203 cm⁻¹ | 275 cm⁻¹ | 321 cm⁻¹ | 428 cm⁻¹ | 561 cm⁻¹ |
| BFCO2 | 155 cm⁻¹ | 199 cm⁻¹ | 274 cm⁻¹ | 324 cm⁻¹ | 429 cm⁻¹ | 561 cm⁻¹ |

Table 3: Variation of activation energy (calculated from imaginary part of impedance and modulus), ferroelectric and magnetic properties with increasing x concentration in BFO.

| Samples | Activation energy (eV) | | $P_s$ ($\mu C/cm^2$) | $2P_r$ ($\mu C/cm^2$) | $2E_c$ (kV/cm) | $M_r$ (emu/gm) | $H_c$ (Oe) |
|---------|------|------|--------|--------|--------|--------|-----|
| | Z'' | M'' | | | | | |
| BFO | 0.91 | 0.87 | 0.2486 | 0.1193 | 11.451 | 0.0007 | 42 |
| BFCO0.5 | 0.93 | 0.89 | 0.2697 | 0.1614 | 15.165 | 0.0102 | 226 |
| BFCO1 | 0.94 | 0.92 | 0.2901 | 0.1999 | 17.181 | 0.0721 | 264 |
| BFCO1.5 | 0.96 | 0.93 | 0.3112 | 0.2421 | 19.239 | 0.1041 | 260 |
| BFCO2 | 0.97 | 0.95 | 0.3386 | 0.2776 | 20.697 | 0.1401 | 296 |

**Figures**

**Fig. 1**

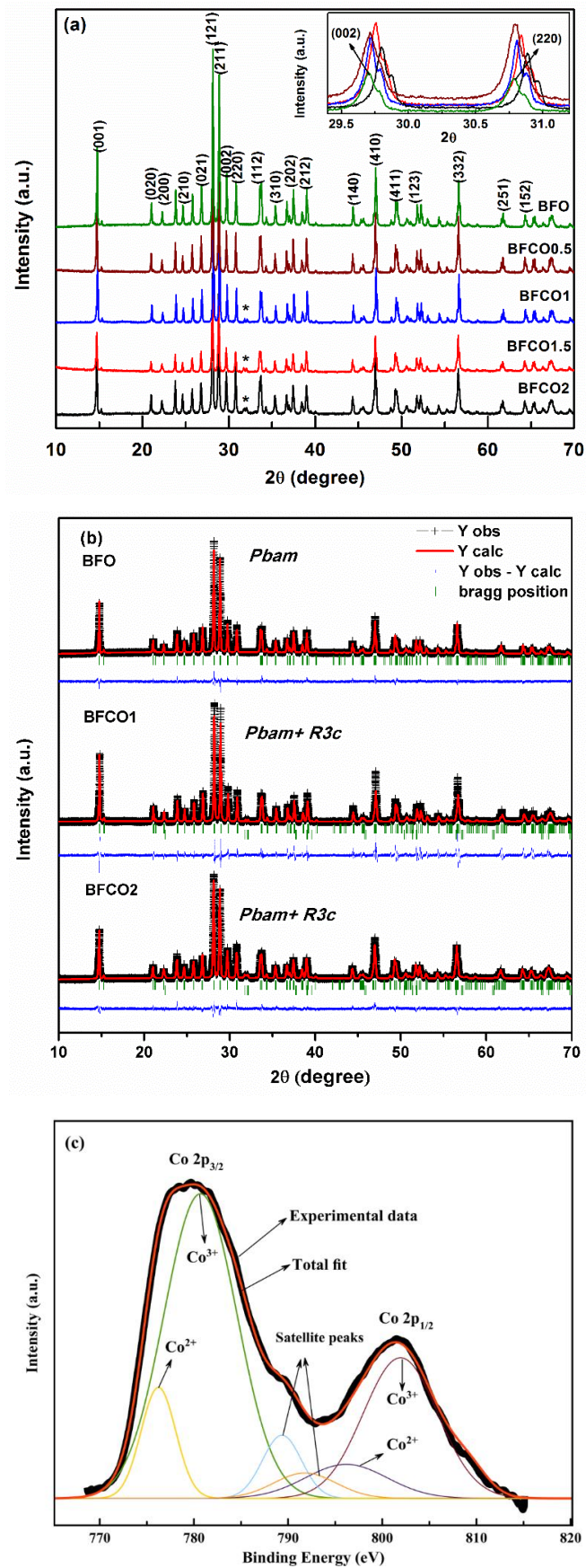

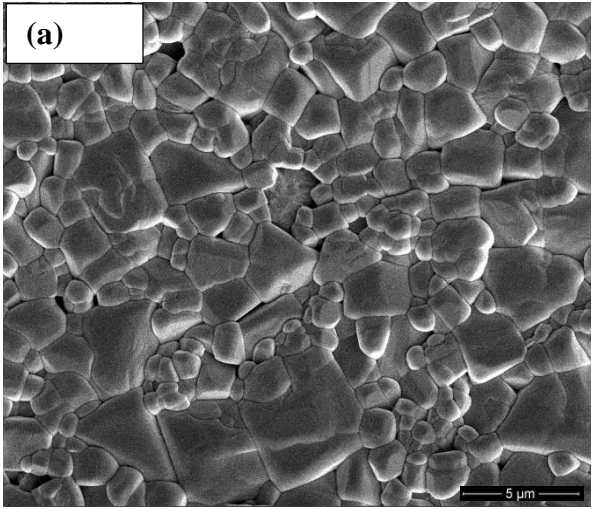 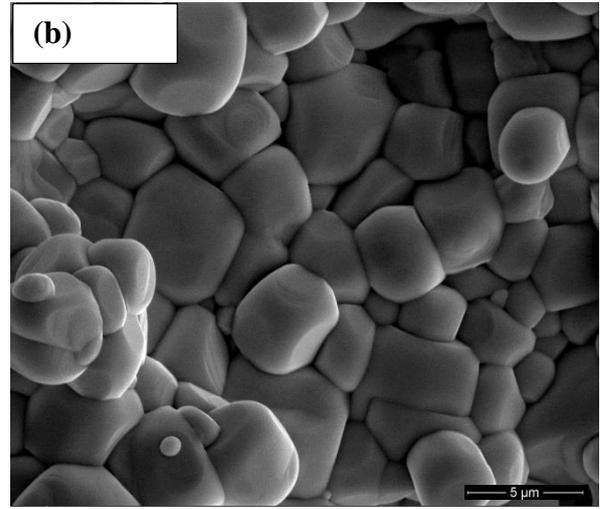

**Fig. 3**

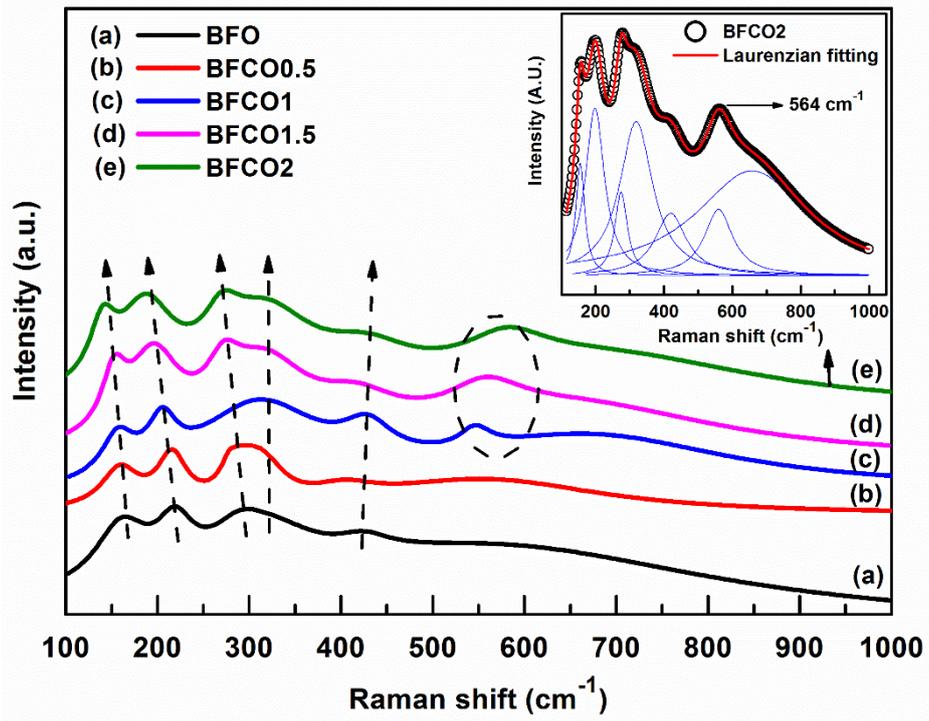

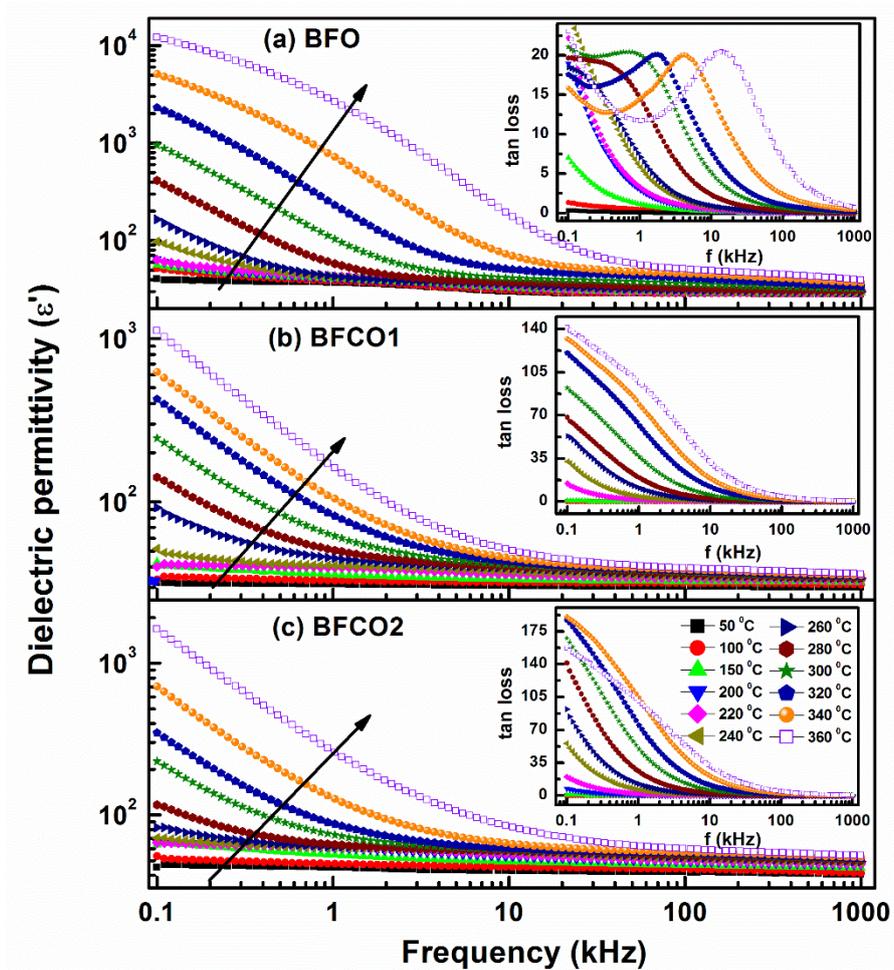

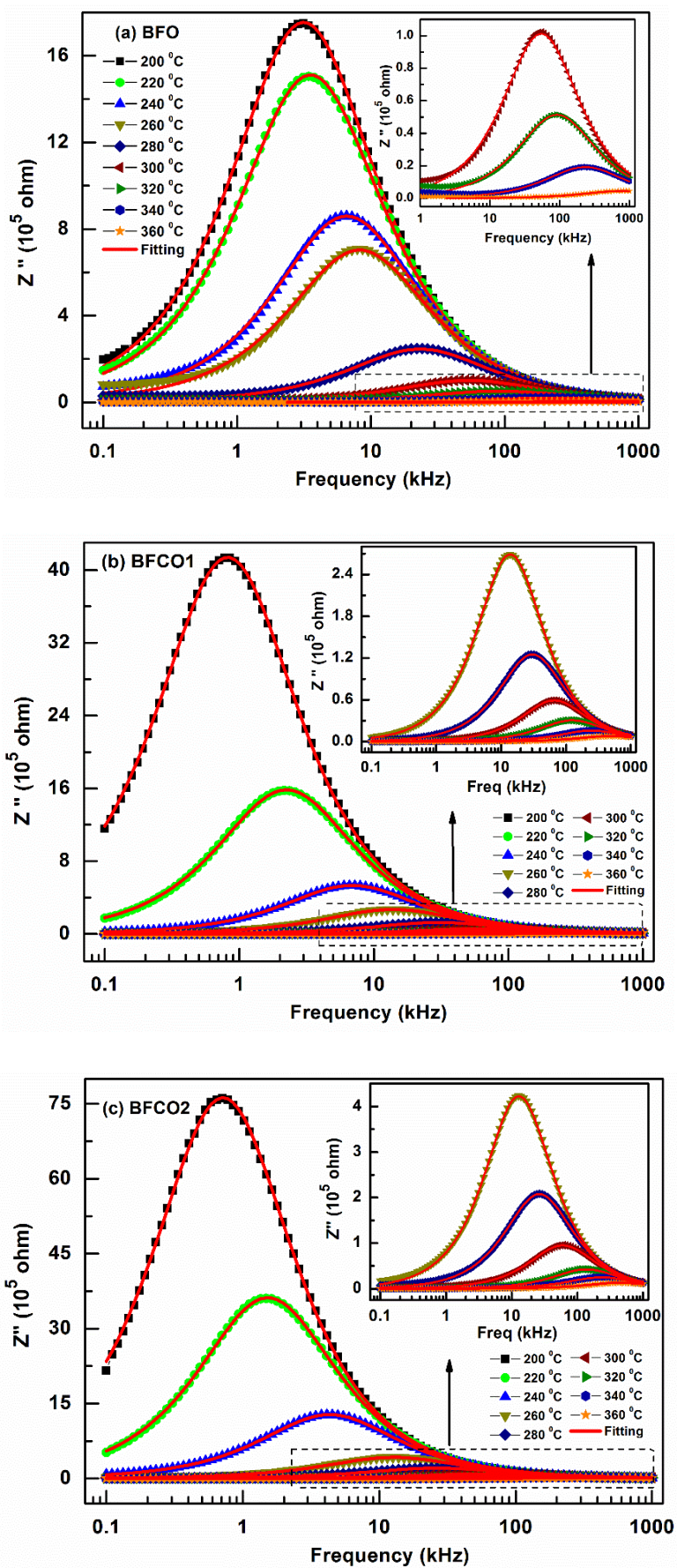

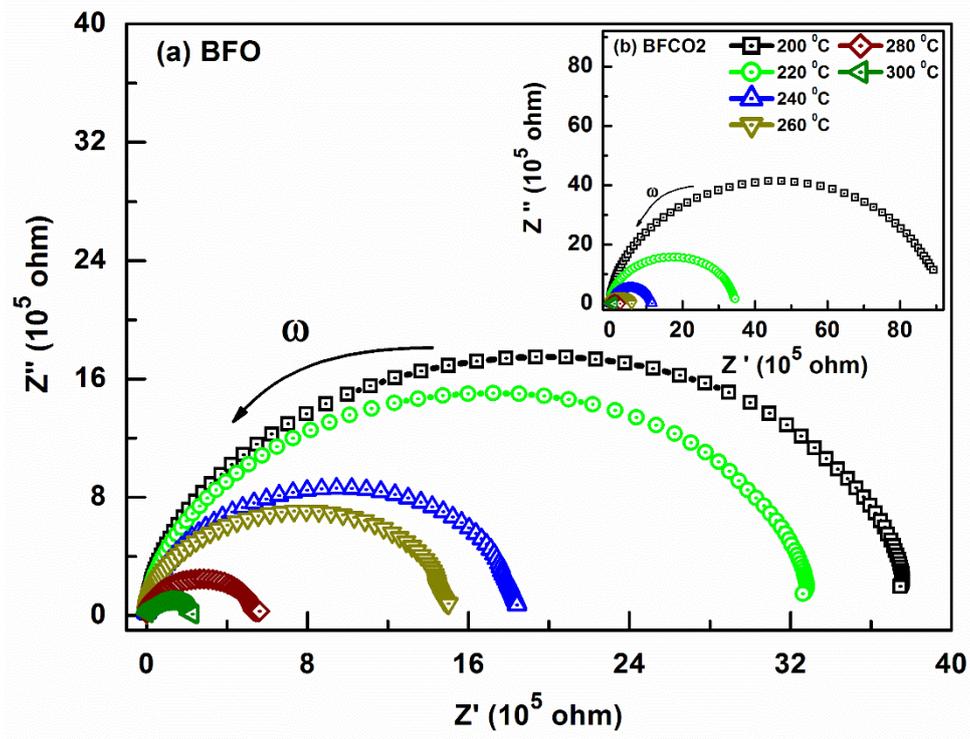

**Fig. 7**

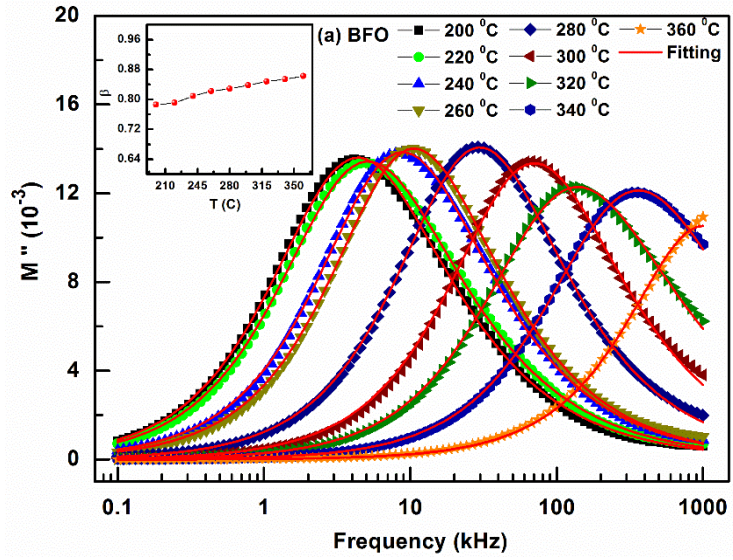

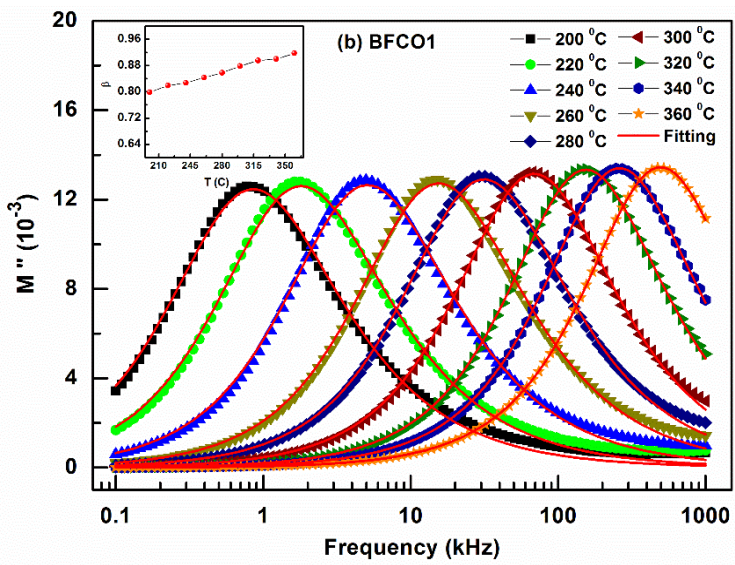

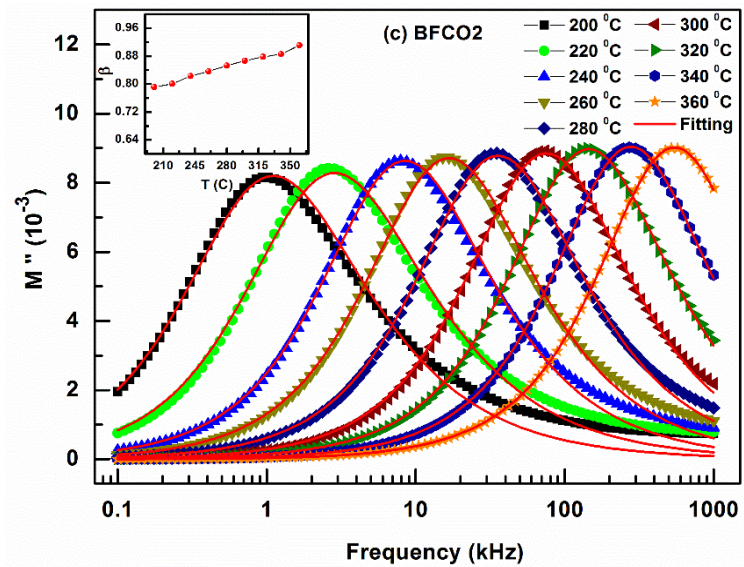

**Fig. 8**

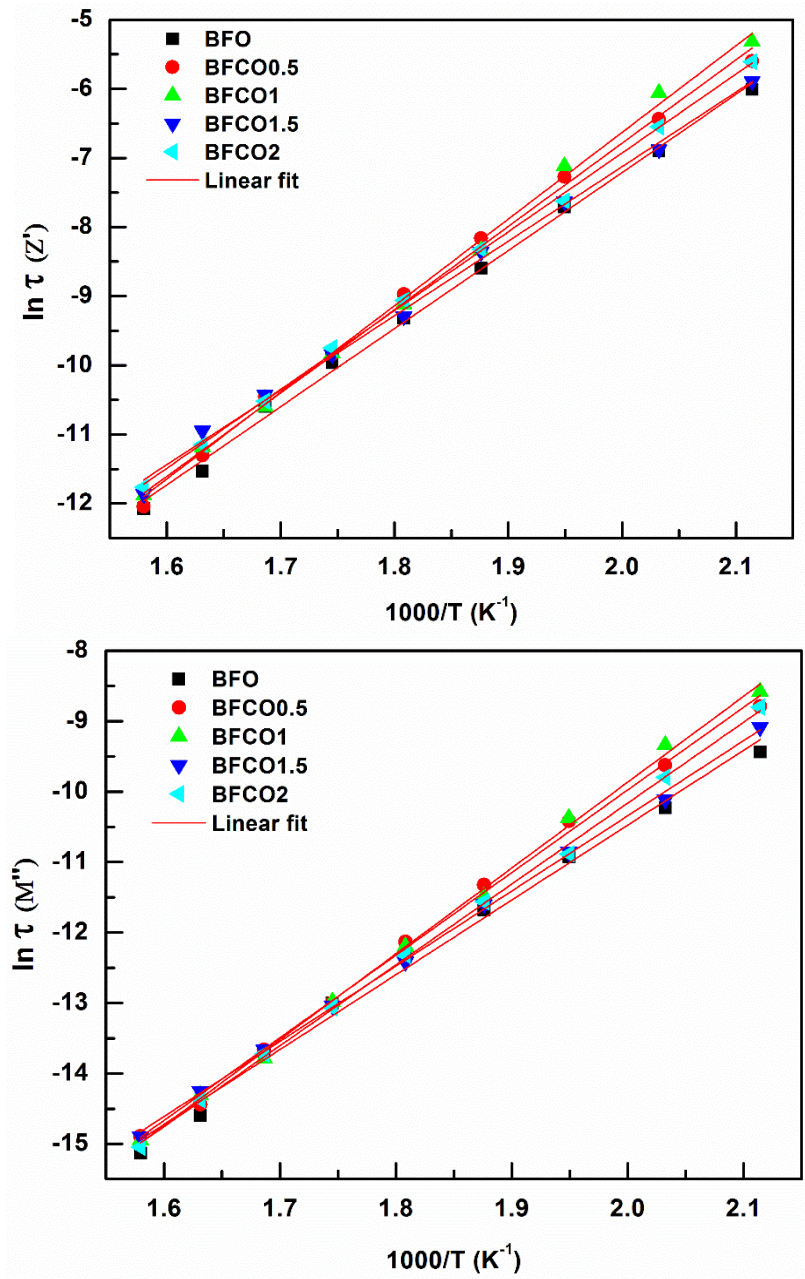

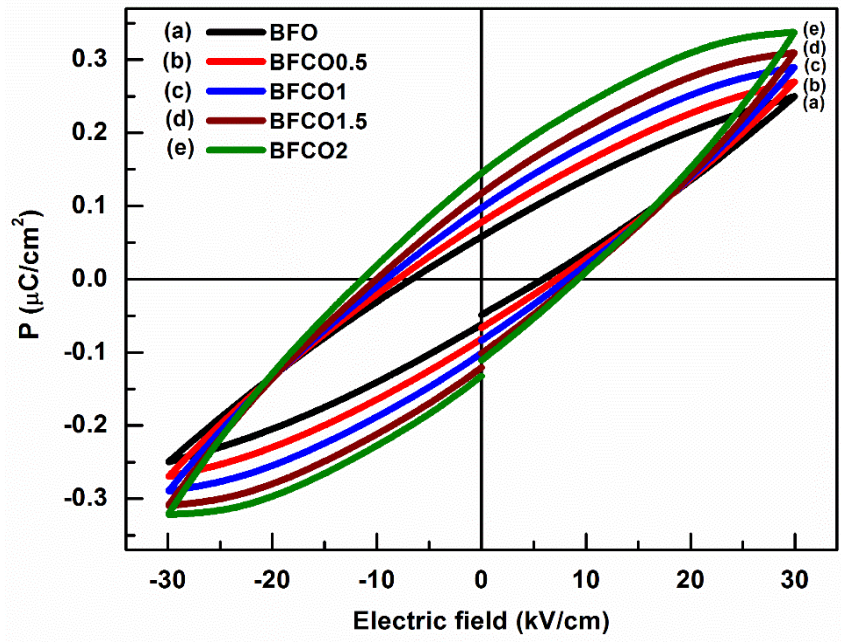

**Fig. 10**

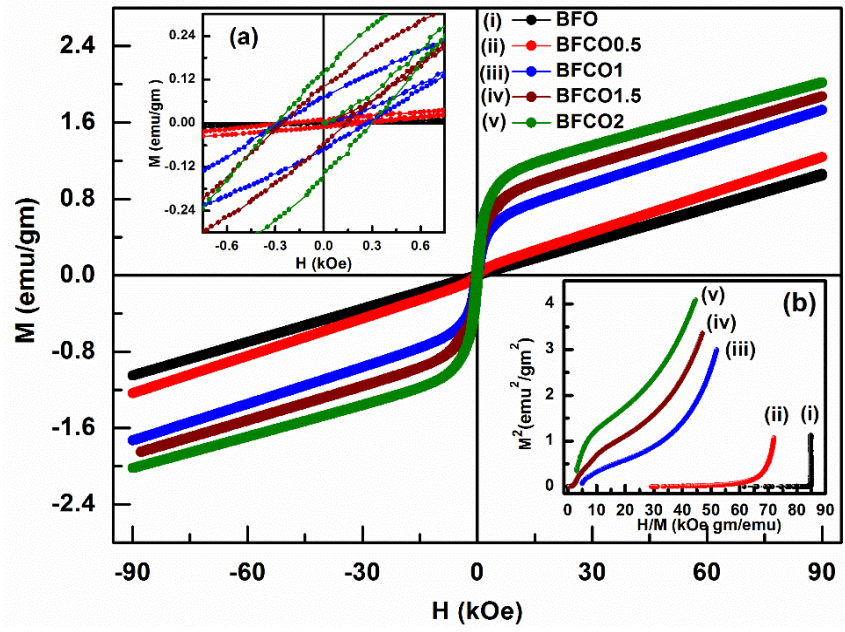